\documentclass[apj]{emulateapj}
\usepackage{times}\usepackage{mathptmx}
\usepackage{xcolor}

\slugcomment{Accepted by ApJ for publication}

\begin{document}

\title{NLTE Analysis of High Resolution {\it{H}}-band Spectra. \\II. Neutral Magnesium \altaffilmark{$\ast$}}
\altaffiltext{$\ast$}{Based on observations collected on the 2.16m telescope at Xinglong station, National Astronomical Observatories, Chinese Academy of Sciences, the 2.2m telescope at the Calar Alto Observatory, the 1.88m reflector on the Okayama Astrophysical Observatory, the Kitt Peak coud\'e feed telescope, and the McMath-Pierce solar telescope and the coud\'e focus of the Mayall 4m reflector at Kitt Peak.}

\author{Junbo Zhang\altaffilmark{1,2}, Jianrong Shi\altaffilmark{1,2}, Kaike Pan\altaffilmark{3}, Carlos Allende Prieto\altaffilmark{4,5}, Chao Liu\altaffilmark{1}}
\affil{\altaffilmark{1} Key Laboratory of Optical Astronomy, National Astronomical Observatories, Chinese Academy of Sciences, A20 Datun Road, Chaoyang District, Beijing 100012, China}
\affil{\altaffilmark{2} University of Chinese Academy of Sciences, Beijing 100049, China}
\affil{\altaffilmark{3} Apache Point Observatory and New Mexico State University, P.O. Box 59, Sunspot, NM, 88349-0059, USA}
\affil{\altaffilmark{4} Instituto de Astrof\'isica de Canarias, 38205 La Laguna, Tenerife, Spain}
\affil{\altaffilmark{5} Departamento de Astrof\'isica, Universidad de La Laguna, 38206 La Laguna, Tenerife, Spain}
\email{sjr@bao.ac.cn}

\begin{abstract}
Aiming at testing the validity of our magnesium atomic model and investigating the effects of non-local thermodynamical equilibrium (NLTE) on the formation of the {\it{H}}-band neutral magnesium lines, we derive the differential Mg abundances from selected transitions for 13 stars either adopting or relaxing the assumption of local thermodynamical equilibrium (LTE). Our analysis is based on high-resolution and high signal-to-noise ratio {\it{H}}-band spectra from the Apache Point Observatory Galactic Evolution Experiment (APOGEE) and optical spectra from several instruments. The absolute differences between the Mg abundances derived from the two wavelength bands are always less than 0.1\,dex in the NLTE analysis, while they are slightly larger for the LTE case. This suggests that our Mg atomic model is appropriate for investigating the NLTE formation of the {\it{H}}-band Mg lines. The NLTE corrections for the \ion{Mg}{1} {\it{H}}-band lines are sensitive to the surface gravity, becoming larger for smaller log $g$ values, and strong lines are more susceptible to departures from LTE. For cool giants, NLTE corrections tend to be negative, and for the strong line at 15765\,\AA\ they reach $-$0.14\,dex in our sample, and up to $-$0.22\,dex for other APOGEE stars. Our results suggest that it is important to include NLTE corrections in determining Mg abundances from the {\it{H}}-band \ion{Mg}{1} transitions, especially when strong lines are used.
\end{abstract}


\section{INTRODUCTION}
Magnesium is a key element in the universe, and plays a significant part in various astrophysical applications. As a typical $\alpha$-element, magnesium is mostly produced by Type II supernovae \citep{woo95}. Thus, it is a good tracer to study $\alpha$-process nucleosynthesis. Unlike Fe, prone to be affected by Type-Ia SNae nucleosynthesis, magnesium is a reliable reference element for the early evolution of the Milky Way \citep{zha98}. \citet{shi98} and \citet{and10} recommended Mg instead of iron as a reference element to investigate the evolution of different abundance ratios, while Si and Ca are more easily influenced by mixing and fallback episodes and possible contributions. \citet{fuh98,fuh04} applied [Mg/Fe] as a reference to explore thick- and thin-disk stars, and found there is a distinct behavior between the two populations. Next to Fe, Mg is an important donor of free electrons in relatively
cool turnoff stellar atmospheres, and contributes significant ultraviolet (UV) opacity \citep{kur79,mas13}. Furthermore, it is one of the best observed elements, inasmuch as several strong Mg lines are easily observed in visible and infrared regions in A to late-type stars.

In the last decades, many studies have determined LTE Mg abundances and have used this element as a tracer to probe the chemical evolution history of our Galaxy, e.g. \citet{lam78}, \citet{tom85}, \citet{che00}, \citet{red06}, \citet{ben14}. However, it is known that Mg abundances may be impacted by NLTE effects. At the end of 1960s, \citet{ath69} first included NLTE effects in the analysis of \ion{Mg}{1} b lines in the Sun based on a small atomic model. Many subsequent studies have focused on investigating the NLTE effects on \ion{Mg}{1} line formation, e.g. \citet{lem87,mau88,gig88,cha91,car92}. \citet{zha98} and \citet{zha00} carefully investigated NLTE effects for \ion{Mg}{1} lines in the Sun and ten cool stars, and found the abundance corrections in the Sun to be negligible, while they can reach $\sim$ 0.1\,dex in metal-deficient stars. They found that the corrections increase with decreasing metallicity, which was confirmed by \citet{geh06}. The latter authors derived NLTE abundances of Na, Mg and Al for a sample of 55 nearby metal-poor stars, and concluded that the ratio of [Al/Mg] is a promising discriminant between thick-disk and halo stellar populations. For extremely metal-poor stars, \citet{and10} studied the NLTE line formation for Mg, and they argued that NLTE effects could, at least partly, explain the unexpected scatter of [X/Mg] found by \citet{cay04} and \citet{bon09}, and the different behaviors between dwarfs and giants described in \citet{bon09}.

Recently, \citet{mas13} improved the Mg atomic model by introducing the newly calculated inelastic collisions with neutral hydrogen from \citet{bar12}, which were based on quantum mechanical computations, and verified that the updated atomic data can improve the determinations of Mg abundances for late-type stars. Her results were confirmed by \citet{oso15}, who produced a novel model atom of Mg and tested its validity for spectral line formation in late-type stars. They predicted that NLTE effects for solar type and metal-poor dwarfs are even smaller than those found in previous studies. However, NLTE corrections can reach up to 0.4\,dex for giants. \citet{ber15} carried out a NLTE analysis of near-infrared $J$-band \ion{Mg}{1} lines for red supergiants.Their results show that NLTE corrections are substantial in the atmospheres of red supergiants, and vary smoothly between $-$0.4\,dex and $-$0.1\,dex as a function of the effective temperature.

The APOGEE survey\footnote{http://www.sdss.org/surveys/apogee}, part of the Sloan Digital Sky Survey III (SDSS-III)\footnote{http://www.sdss3.org} \citep{eis11}, has observed $\sim$150,000 predominantly red giants stars covering the full range of Galactic bulge, bar, disk and halo \citep{maj15} since 2011, and APOGEE-2, an on-going extension of the project in SDSS-IV will significantly enlarge this data base. APOGEE infrared {\it{H}}-band spectra have already been publicly released as part of the SDSS Data Release 10 (DR10) \citep{ahn14} and Data Release 12 (DR12) \citep{ala15}. These observations provide a promising way to trace and to explore the formation history of the Milky Way. 

The APOGEE Stellar Parameters and Chemical Abundances Pipeline (ASPCAP) \citep{gar16} is designed to derive the stellar parameters (effective temperature, surface gravity, metallicity), and chemical abundances of 15 different elements. As pointed out by \citet{mes13}, NLTE effects may have an impact on the APOGEE derived stellar parameters and chemical abundances due to low densities present in the atmospheres of giants.  
Thus, as an extension of our previous work \defcitealias{zha16}{Paper~I}\citepalias[hereinafter,][]{zha16}, which is confined to {\it{H}}-band Si lines, this paper focuses on NLTE line formation of \ion{Mg}{1} {\it{H}}-band transitions for the 13 sample stars from \citetalias{zha16}. The main purpose of this work is to validate the applied Mg atomic model, and to investigate the influence of departures from LTE. This will help us to improve the accuracy of stellar parameters determined by the APOGEE pipeline and help to explore the chemical enrichment history of the Galaxy based on large samples of APOGEE {\it{H}}-band spectra.

This paper is organized as follows. In Section\,\ref{method}, we introduce the Mg model atom and the NLTE calculations. The data and the determination of stellar parameters for our sample stars are briefly described in Section\,\ref{sample_access}. Section\,\ref{nlte} derives the Mg abundances from both {\it{H}}-band and optical lines for our sample stars under LTE and NLTE analyses, respectively, and compares derived Mg abundances from the two bands, discussing the implications. In the last section, we summarize our results.

\section{METHOD OF NLTE CALCULATIONS} \label{method}
\subsection{Model Atom of Magnesium} \label{atom}
We utilized the updated model atom for Mg from \citet{mas13}, which was based on the model produced by \citet{zha98} and \cite{zha00}. This comprehensive model atom contains the first three ionization stages of Mg, including 85 terms of \ion{Mg}{1}, two levels of \ion{Mg}{2} and the ground state of \ion{Mg}{3}. Here, we briefly introduce the atomic data. For electron impact excitation, the collisional rates were taken from \citet{mau88} when available, while from \citet{zha98} for the rest of the transitions. Ionization cross-sections were computed from the formula by \citet{sea62}. For hydrogen-impact excitation and charge transfer, the rate coefficients of the transitions between the seven lowest levels of Mg and the ionic state (Mg $+$ H and Mg$^+$ $+$ H$^-$) were adopted from \citet{bar12} \citep[see][for details]{mas13}. 

\subsection{Model Atmospheres} \label{model}
In this study, we adopted the widely-used grid of MARCS atmosphere models\footnote{http://marcs.astro.uu.se} \citep{gus08}. These LTE models are divided in two groups: models with 3.0 $\leq$ log $g \leq 5.5$ were computed with a plane-parallel geometry, while those with relatively low surface gravities (-1.0 $\leq$ log $g$ $\leq$ 3.5) were calculated in spherical geometry. As in \citetalias{zha16}, according to the suggestions by \citet{gus08} and \citet{hei06}, spherical models were adopted for stars with log $g$ $\le$ 3.5 and plane-parallel model atmospheres for the rest. The final models were derived by interpolating with a FORTRAN routine coded by Thomas Masseron\footnote{http://marcs.astro.uu.se/software.php}. MARCS model atmospheres \citep{gus08} adopt a solar chemical composition from \citet{gre07}. An $\alpha$-enhancement is considered, and the mixing-length parameter is set to $l/\rm{H}_P$=1.5. 

\subsection{Statistical Equilibrium Codes} \label{codes}
As in \citetalias{zha16}, we adopted a revised version of the DETAIL code \cite[]{but85} to solve the coupled statistical equilibrium and the radiative transfer equations. This statistical equilibrium code is based on the accelerated lambda iteration algorithm described by \citet{ryb91,ryb92}, which has been widely applied in previous studies, e.g. \citet{geh04}, \citet{shi08}, \citet{mas11,mas13}, \citet{ber12} and \citet{sit15}, etc. We calculated the departure coefficients using DETAIL, and then input these coefficients to SIU \citep{ree91} to calculate the synthetic NLTE line profiles.

\section{THE SAMPLE STARS AND STELLAR PARAMETERS} \label{sample_access}
\subsection{Sample Selection } \label{sample}
We selected 13 stars with both IR {\it{H}}-band and optical high-resolution and high S/N spectra available as sample stars. The {\it{H}}-band spectra are mainly from APOGEE DR12 except those for the Sun and \object{Arcturus}, which will be described below. These APOGEE data were acquired with the NMSU (new Mexico State University) 1m telescope coupled to the APOGEE instrument: a bundle of ten fibers connects the APOGEE spectrograph and the NMSU 1m telescope. In each observation, one fiber is assigned to the science target while the rest are used as sky fibers. (See \citet{feu16} for more details.) The wavelength range of APOGEE spectra spans from 15100 to 16900\,\AA, and the resolving power is $\sim$ 22,500. A high-resolution and high S/N {\it{H}}-band spectrum of the Sun was taken with the McMath-Pierce solar telescope on Kitt Peak \citep{wal96}, and corresponds to the disk center region ($\mu = 1$). A high quality IR spectrum of \object{Arcturus} was obtained from the NOAO data archives\footnote{http://ast.noao.edu/data/} (see \citet{hin95} for more observational information). The resolving power is around 300,000 and 100,000, respectively, for the Sun and \object{Arcturus}. The high-resolution optical spectra of the 13 sample stars are from several different telescopes, and their characteristics have been described in detail in \citetalias{zha16}.

\subsection{Stellar parameters} \label{para}
We determined the stellar parameters $T_{\rm{eff}}$, log $g$, [Fe/H] and $\xi_{t}$  using a spectroscopic approach from the analysis of Fe I and Fe II lines. The final parameters satisfy the Fe I excitation equilibrium ($T_{\rm{eff}}$), the ionization equilibrium between Fe I and Fe II (log $g$), and [Fe/H] does not depend on equivalent width ($\xi_t$). NLTE effects on \ion{Fe}{1} lines have been considered as in \citet{sit15}. We estimated the typical uncertainties of $T_{\rm{eff}}$, log $g$, [Fe/H], and $\xi_t$ as $\pm$80\,K, $\pm$0.1\,dex, $\pm$0.08\,dex, and 0.2\,km\,s$^{-1}$, respectively. The determination of the stellar parameters and the comparison with previous work have been discussed in \citetalias{zha16}.

\section{NLTE CALCULATIONS FOR SAMPLE STARS} \label{nlte}
\subsection{Line Data} \label{line}
\subsubsection{Infrared Atomic Line Data in the {\it{H}}-band} \label{h_line}
Magnesium abundances were derived from eight {\it{H}}-band \ion{Mg}{1} lines, and the characteristics of these transitions are listed in Table\,\ref{tbl-1}. The damping constants, log $C_6$, are adopted from \citet{mel99}, which were calculated according to the {\it{ABO}} theory \citep{ans95,bar97}. The rest of the line data were adopted from the NIST atomic spectra database\footnote{http://www.nist.gov/pml/data/asd.cfm}. The values of oscillator strengths, log $gf$, were rescaled with respect to the solar NLTE results (adopting log $\varepsilon_\sun$(Mg) $=$ 7.53\,dex measured from meteorites according to \citet{gre07} as the absolute solar Mg abundance). Among these eight lines, two transitions at 15748.886 and 15748.988\,\AA, three transitions 
at 15765.645, 15765.747, and 15765.842\,\AA, and transitions at 15886.183 and 15886.261\,\AA \,are severally blended, and they cannot be resolved at the resolution that the APOGEE instrument provides. However, it is worthwhile noting that each of the three blended lines is due to transitions between the same lower (or higher) energy level and fine structure splitting of the higher (or lower) level, we therefore fit all lines in a blended feature together via spectrum synthesis in a small wavelength interval. As a result, only four values for the Mg abundances from the eight lines are actually derived for each star from the infrared spectra.

\begin{deluxetable}{lccrcc}
\tabletypesize{\scriptsize}
\tablecaption{Atomic data of the optical and H-band magnesium lines\label{tbl-1}}
\tablewidth{0pt}
\tablehead{
\colhead{$\lambda$ (\AA)} & \colhead{Transition\tablenotemark{a}} & \colhead{$\chi$ (eV)} & \colhead{log $gf$} & \colhead{log $C_6$}
}
\startdata
4571.096  & 3$s^{\rm{2}}~^{\rm{1}}$S$_{\rm{0}}-$3$p~^{\rm{3}}$P$^{\rm{o}}_{\rm{1}}$ & 0.000 & $-$5.49 & $-$31.799\tablenotemark{b}\\
4702.991  & 3$p~^{\rm{1}}$P$^{\rm{o}}_{\rm{1}}-$5$d~^{\rm{1}}$D$_{\rm{2}}$   & 4.346 & $-$0.36 & $-$29.849\tablenotemark{b}\\
5172.684  & 3$p~^{\rm{3}}$P$^{\rm{o}}_{\rm{1}}-$4$s~^{\rm{3}}$S$_{\rm{1}}$   & 2.712 & $-$0.44 & $-$30.549\tablenotemark{b}\\
5183.604  & 3$p~^{\rm{3}}$P$^{\rm{o}}_{\rm{2}}-$4$s~^{\rm{3}}$S$_{\rm{1}}$   & 2.717 & $-$0.21 & $-$30.549\tablenotemark{b}\\
5528.405  & 3$p~^{\rm{1}}$P$^{\rm{o}}_{\rm{1}}-$4$d~^{\rm{1}}$D$_{\rm{2}}$   & 4.346 & $-$0.40 & $-$30.324\tablenotemark{b}\\
5711.088  & 3$p~^{\rm{1}}$P$^{\rm{o}}_{\rm{1}}-$5$s~^{\rm{1}}$S$_{\rm{0}}$   & 4.346 & $-$1.70 & $-$29.890\tablenotemark{c}\\
\hline
15740.716 & 4$p~^{\rm{3}}$P$^{\rm{o}}_{\rm{0}}$$-$4$d~^{\rm{3}}$D$_{\rm{1}}$ & 5.932 & $-$0.36 & $-$29.658\tablenotemark{d}\\
15748.886 & 4$p~^{\rm{3}}$P$^{\rm{o}}_{\rm{1}}$$-$4$d~^{\rm{3}}$D$_{\rm{1}}$ & 5.932 & $-$0.54 & $-$29.658\tablenotemark{d} \\
15748.988 & 4$p~^{\rm{3}}$P$^{\rm{o}}_{\rm{1}}$$-$4$d~^{\rm{3}}$D$_{\rm{2}}$ & 5.932 &    0.02 & $-$29.658\tablenotemark{d} \\
15765.645 & 4$p~^{\rm{3}}$P$^{\rm{o}}_{\rm{2}}$$-$4$d~^{\rm{3}}$D$_{\rm{1}}$ & 5.933 & $-$1.54 & $-$29.658\tablenotemark{d} \\
15765.747 & 4$p~^{\rm{3}}$P$^{\rm{o}}_{\rm{2}}$$-$4$d~^{\rm{3}}$D$_{\rm{2}}$ & 5.933 & $-$0.55 & $-$29.658\tablenotemark{d} \\
15765.842 & 4$p~^{\rm{3}}$P$^{\rm{o}}_{\rm{2}}$$-$4$d~^{\rm{3}}$D$_{\rm{3}}$ & 5.933 &    0.30 & $-$29.658\tablenotemark{d} \\
15886.183 & 3$d~^{\rm{3}}$D$_{\rm{2}}$$-$5$p~^{\rm{3}}$P$^{\rm{o}}_{\rm{1}}$ & 5.946 & $-$1.71 & $-$29.569\tablenotemark{d}\\
15886.261 & 3$d~^{\rm{3}}$D$_{\rm{1}}$$-$5$p~^{\rm{3}}$P$^{\rm{o}}_{\rm{1}}$ & 5.946 & $-$2.07 & $-$29.569\tablenotemark{d}
\enddata
\tablenotetext{a}{Transition information is from NIST Atomic Spectra Database.}
\tablenotetext{b}{\citet{mas13}}
\tablenotetext{c}{The damping constant was determined by fitting the line wings of the solar spectrum.}
\tablenotetext{d}{\citet{mel99}}
\end{deluxetable}

\subsubsection{Optical Atomic Line Data} \label{op_line}
There are six optical \ion{Mg}{1} lines included in our analysis, and their main characteristics are presented in Table\,\ref{tbl-1}. Similar to the {\it{H}}-band lines, the oscillator strengths have been scaled in order to have the NLTE computed profiles matching the  observations with log $\varepsilon_\sun$(Mg) $=$ 7.53\,dex. We calculated the van der Waals damping constants for the 4571, 5172 and 5183\,\AA\ lines according to Table\,\ref{tbl-1} of \citet{mas13}, who presented van der Waals broadening constants $\Gamma_{6}$ based on the {\it{ABO}} theory. Following her suggestion, we reduced $\Gamma_{6}$ by 0.3 and 0.2\,dex for the \ion{Mg}{1} 4703 and 5528\,\AA\ lines, respectively. For the 5711\,\AA\ line, the $C_6$ value was determined by fitting the wings of the lines in the solar spectrum \citep{geh04}. Actually these three optical \ion{Mg}{1} lines' C6 values (4703, 5528, 5711) were determined by fitting the solar spectrum, because the values from the ABO theory enhance the abundance discrepancy among the different solar Mg lines.

\subsection{NLTE Effects} \label{nlte_effects}
\subsubsection{Departures form LTE for  \ion{Mg}{1} H-band lines} \label{nlte_h}
\begin{figure}
\includegraphics[scale=0.4,keepaspectratio=true,clip=true]{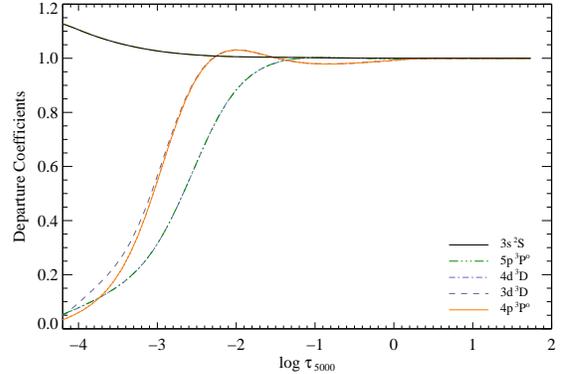}
\caption{Departure coefficients $b_{i} = N^{\rm{NLTE}}_{i}/N_{i}^{\rm{LTE}}$ as a function of the standard optical depth for HD 58367. 4$p~^{\rm3}{\rm{P}}^{\rm{o}}$ (the red solid line) and 3$d~^{\rm3}{\rm{D}}$ (the dashed line) couple with each other. \label{fig1}}
\end{figure}

We present the departure coefficients ($b_{i}$) for the relevant \ion{Mg}{1} levels, including    \ion{Mg}{2} ground state, as a function of the optical depth at 5000\,\AA\ ($\tau_{5000}$) for a model atmosphere for \object{HD\,58367} in Figure \ref{fig1}. The departure coefficient is defined as $b_{i} = n_{i}^{\rm{NLTE}}/n_{i}^{\rm{LTE}}$, while $n_{i}^{\rm{NLTE}}$ and $n_{i}^{\rm{LTE}}$, respectively, indicate NLTE and LTE atomic level number densities. We noted that the populations of the two relative low levels 4$p~^{\rm3}{\rm{P}}^{\rm{o}}$ and 3$d~^{\rm3}{\rm{D}}$ slightly deviate from their LTE values at optical depths near unity. This reduction   in the level population is the result of the large photoionization rate which is known to dominate the near-UV spectra of cool stars. The other two higher excitation levels (4$d~^{\rm{3}}{\rm{D}}$, 5$P~^{\rm3}{\rm{P}}^{\rm{o}}$), are underpopulated due to photon loss at optical depths around $-2$ (see Figure \ref{fig1} for details).

\begin{figure*}
\plottwo{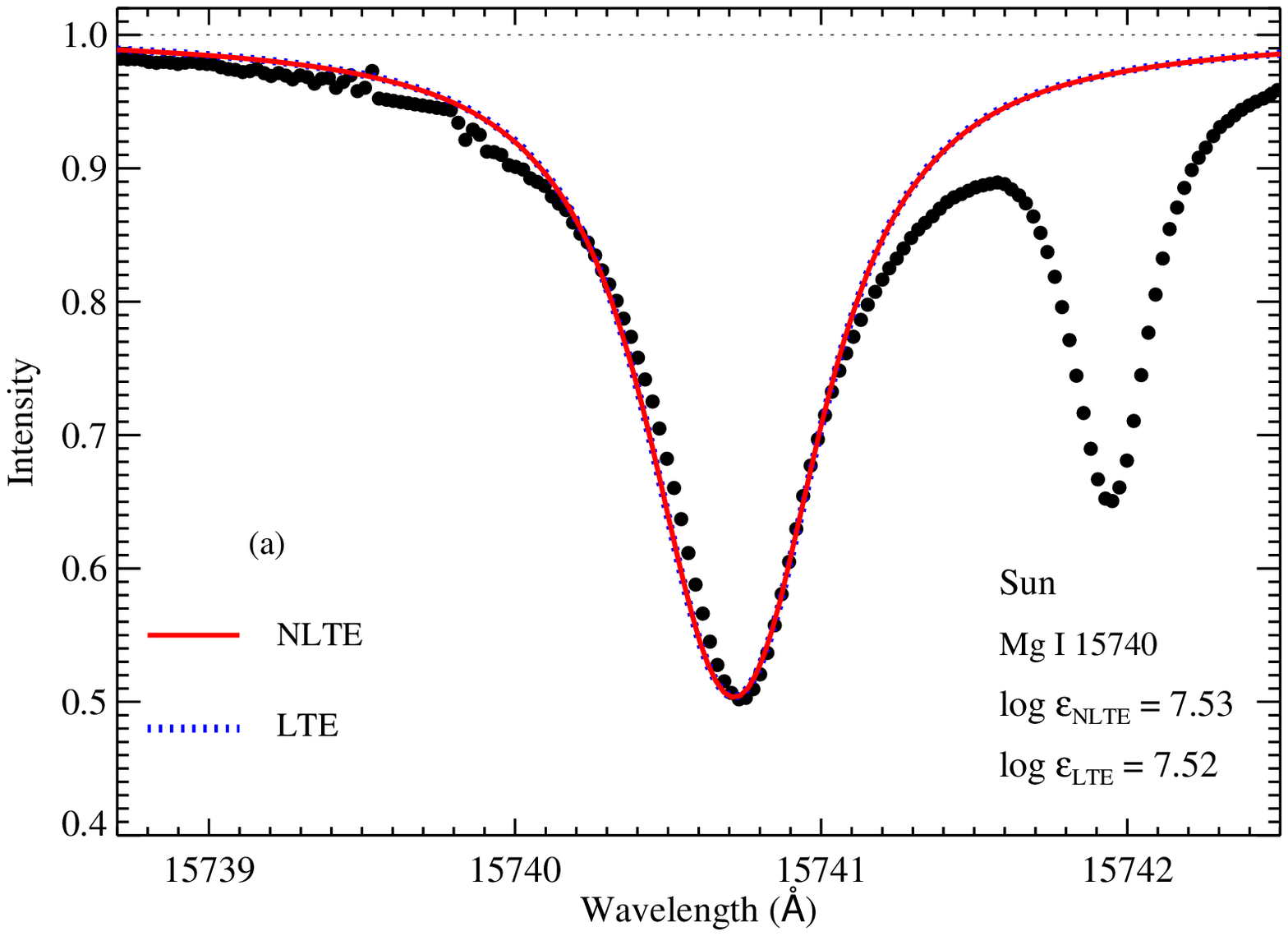}{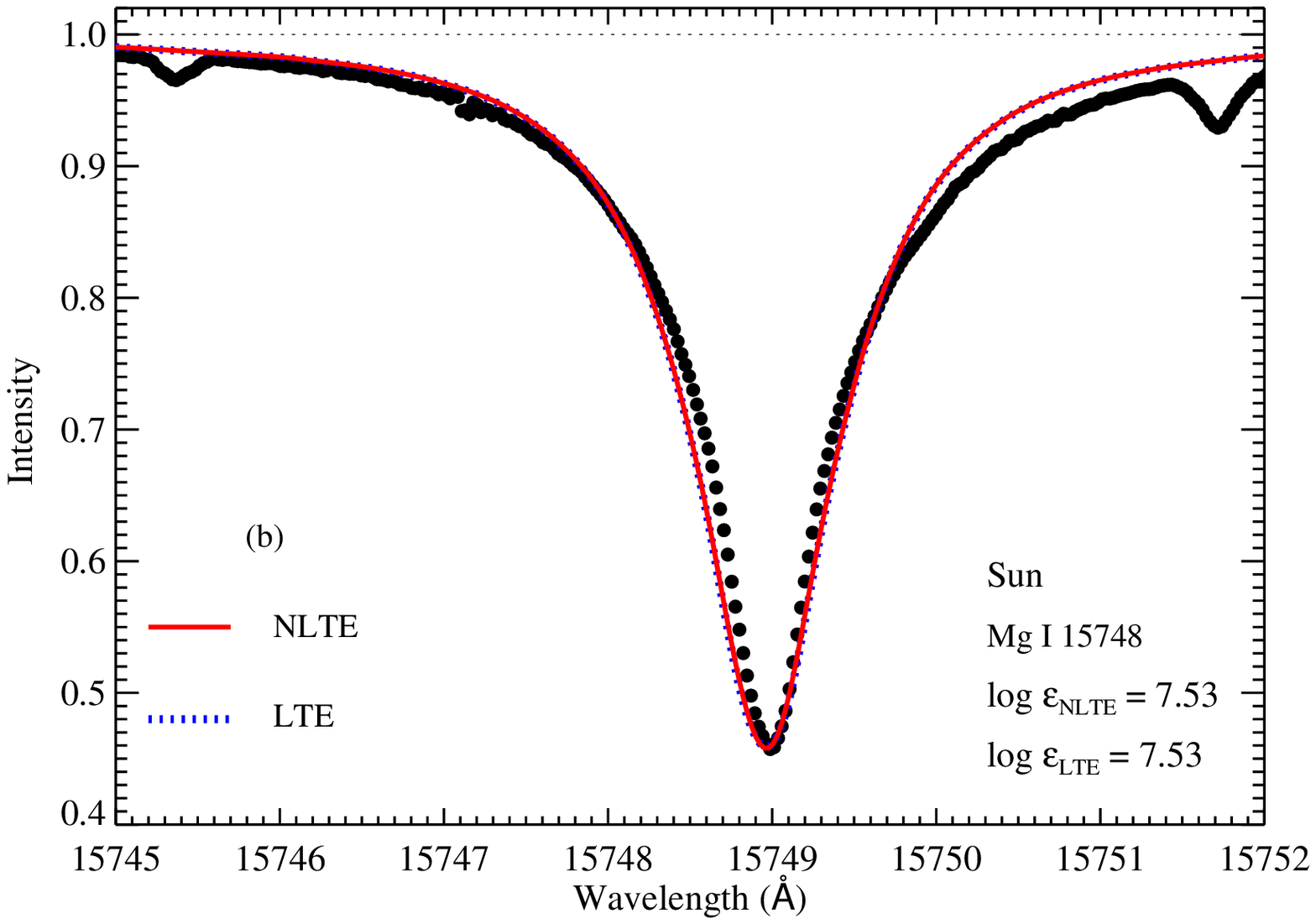}
\plottwo{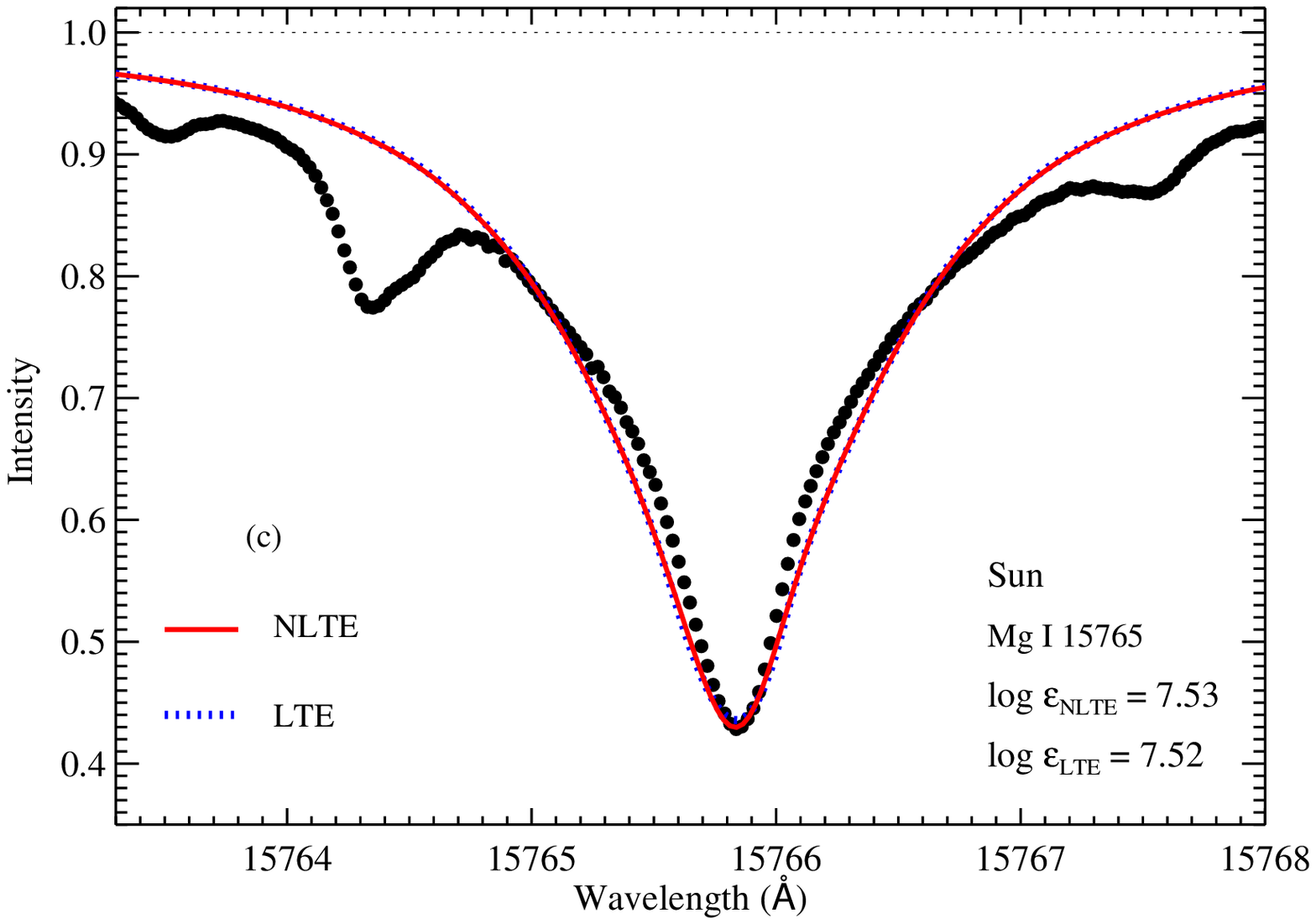}{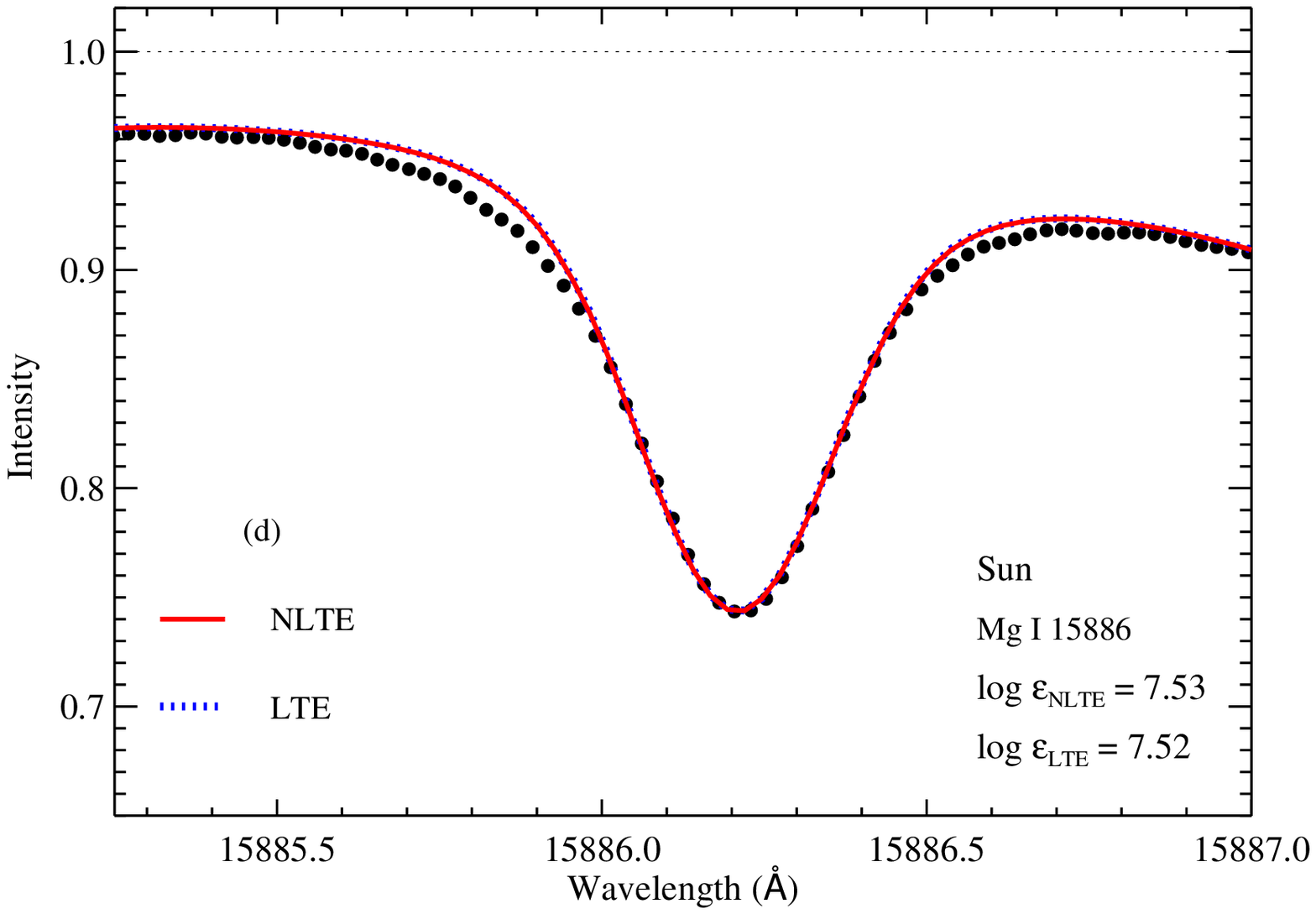}
\caption{Best NLTE (continuous curve) and LTE (dotted curve) fits of the four H-band \ion{Mg}{1} lines in comparison with the observed solar spectrum (filled circles). \label{fig2}}
\end{figure*}

\begin{figure}
\includegraphics[scale=0.45,keepaspectratio=true,clip=true]{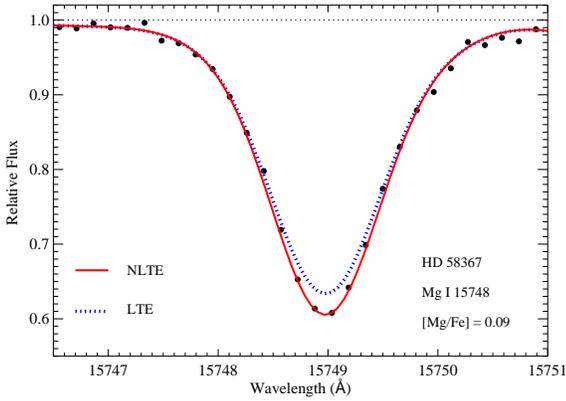}
\caption{The LTE and NLTE synthetic spectra of \ion{Mg}{1} 15748\,\AA\ line with the same [Mg/Fe] for HD\,58367. The dotted curve is the observed spectrum.\label{fig3}}
\end{figure}

The Mg abundances were estimated through spectrum synthesis, and the NLTE effects can be assessed by comparing NLTE and LTE results ($\Delta$ $=$ $\rm{log}\,\varepsilon_{NLTE}$ $-$ $\rm{log}\,\varepsilon_{LTE}$). For each star, spectral synthesis calculations were required to fit the observed line profiles by changing the elemental abundance in the NLTE and LTE cases, respectively. The Mg abundances relative to iron  for our four-group {\it{H}}-band lines are presented in Table\,\ref{tbl-2}. The adopted solar stellar parameters are $T_{\rm{eff}}$ $=$ 5777\,K, [Fe/H] $=$ 0.0\,dex, log $g$ $=$ 4.44\,dex. $\xi_t$ $=$ 0.9\,km\,s$^{-1}$. The {\it{H}}-band solar \ion{Mg}{1} line profiles are computed for $\mu = 1$, and the comparison of the best-fit NLTE synthetic profiles (solid line) and LTE (dotted line) with the observed solar ones (open circles) for the four \ion{Mg}{1} lines is illustrated in Figure \ref{fig2}. We noted that the synthesized line profiles are slightly broad in the central part compared to the observed ones for strong lines, especially for lines at 15748\,\AA\ and 15765\,\AA, as apparent in Figure \ref{fig2}, while the weak line at 15886\,\AA\ is reproduced much better. The systematic deviations between model and observed line profiles for the transitions 15748 and 15765\,\AA\ are apparent both in LTE and NLTE, since NLTE corrections are very small. In addition, the observations for these lines exhibit significant asymmetries between the blue and red wings.  We have tested the influence of micro-turbulence 
and scattering and they have only a minor impact on the computed profiles. These strong lines are sampling a very large range in optical depth, and the noted discrepancies are likely reflecting modeling shortcomings such as atmospheric inhomogeneities in temperature and velocity, or systematic errors in the  line formation associated to them. This also possibly results from an imperfectness of line broadening theory. As mentioned above, the NLTE corrections for the four solar {\it{H}}-band lines are very small, within 0.01\,dex.
Figure \ref{fig3} shows the synthetic flux profiles under LTE (dotted curve) and NLTE (solid curve) assumptions with the same Mg abundance for the line at 15748\,\AA\ in the spectrum of \object{HD\,58367}, and the difference is obvious.

\begin{figure}
\includegraphics[scale=0.45,keepaspectratio=true,clip=true]{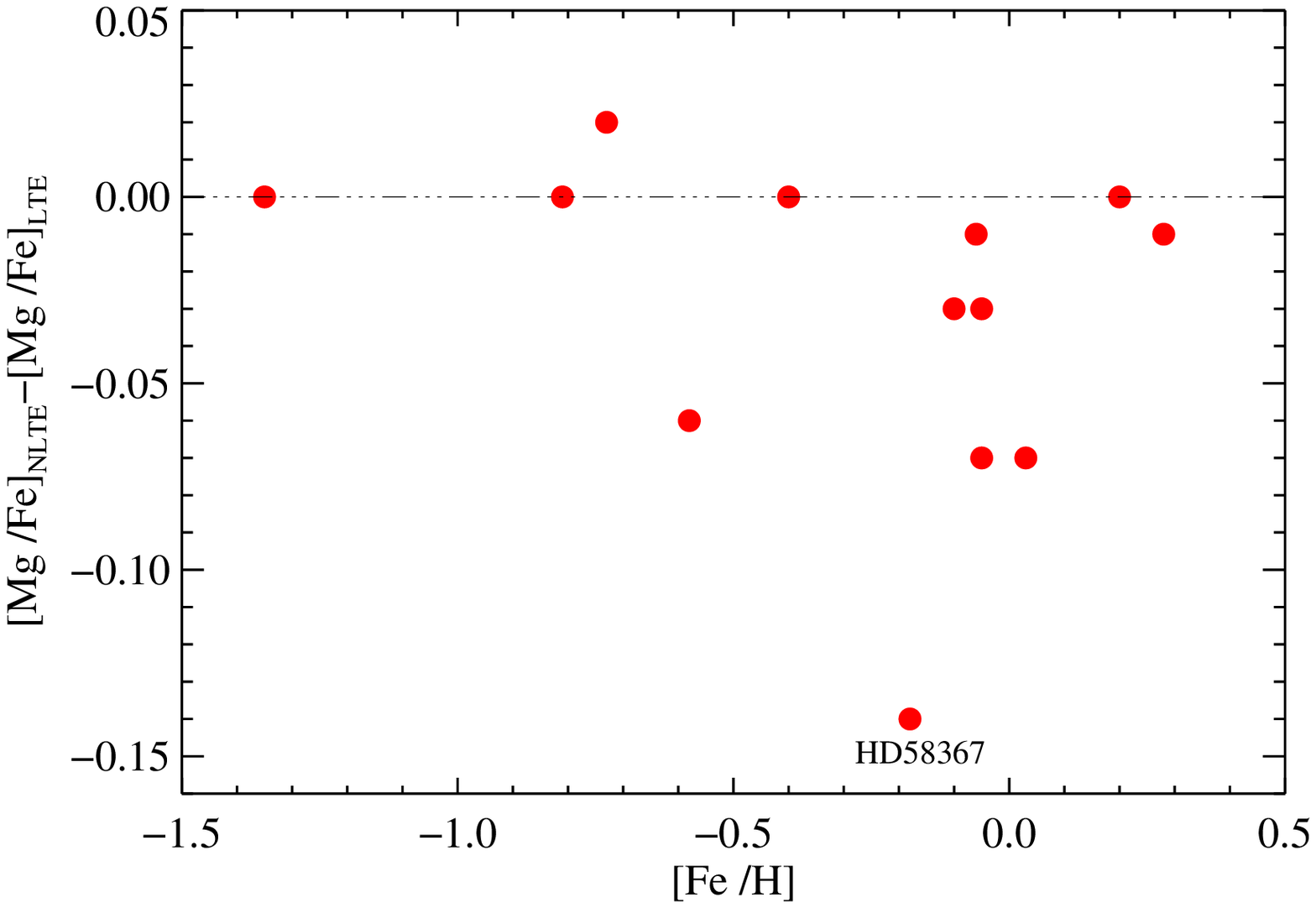}
\includegraphics[scale=0.45,keepaspectratio=true,clip=true]{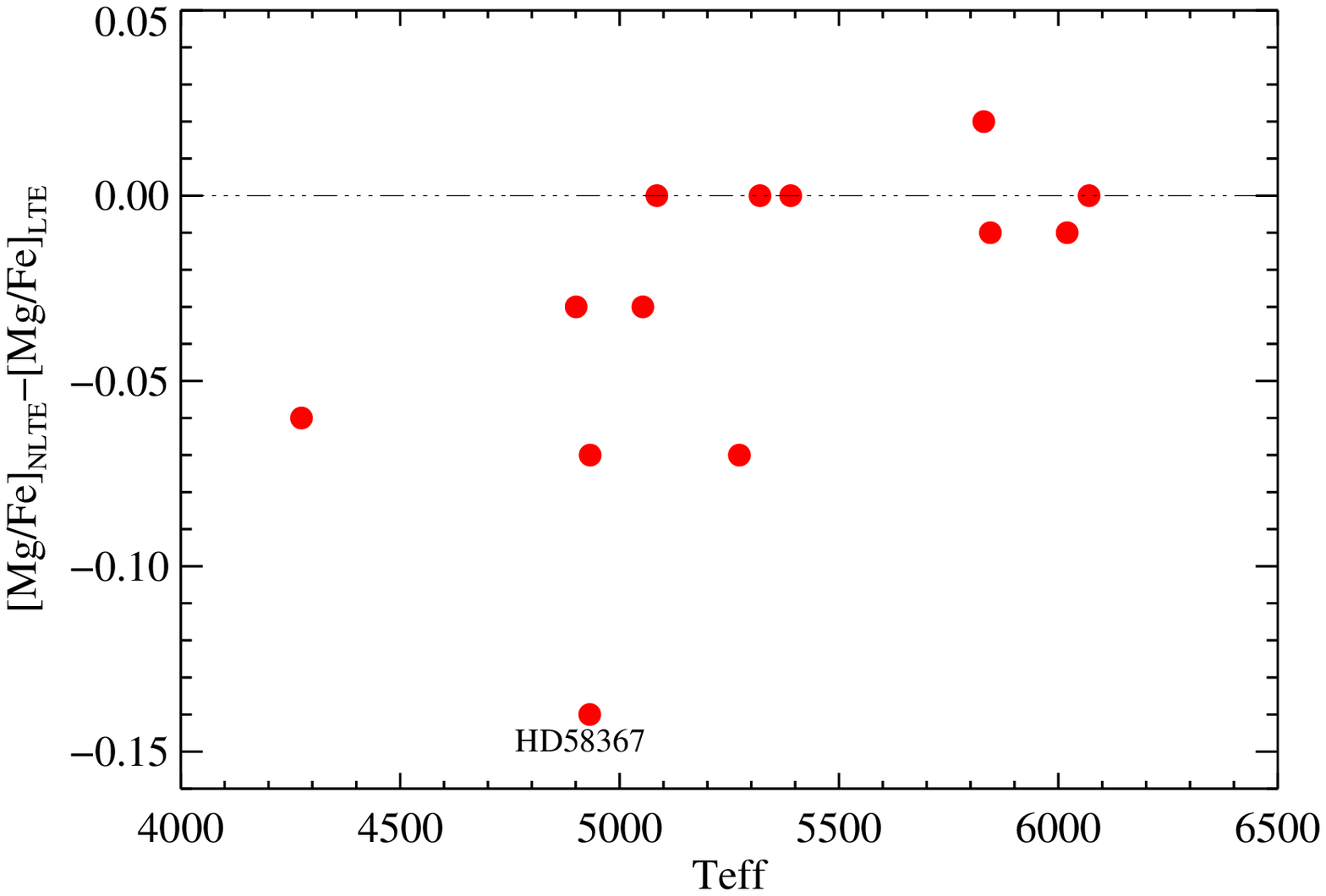}
\includegraphics[scale=0.45,keepaspectratio=true,clip=true]{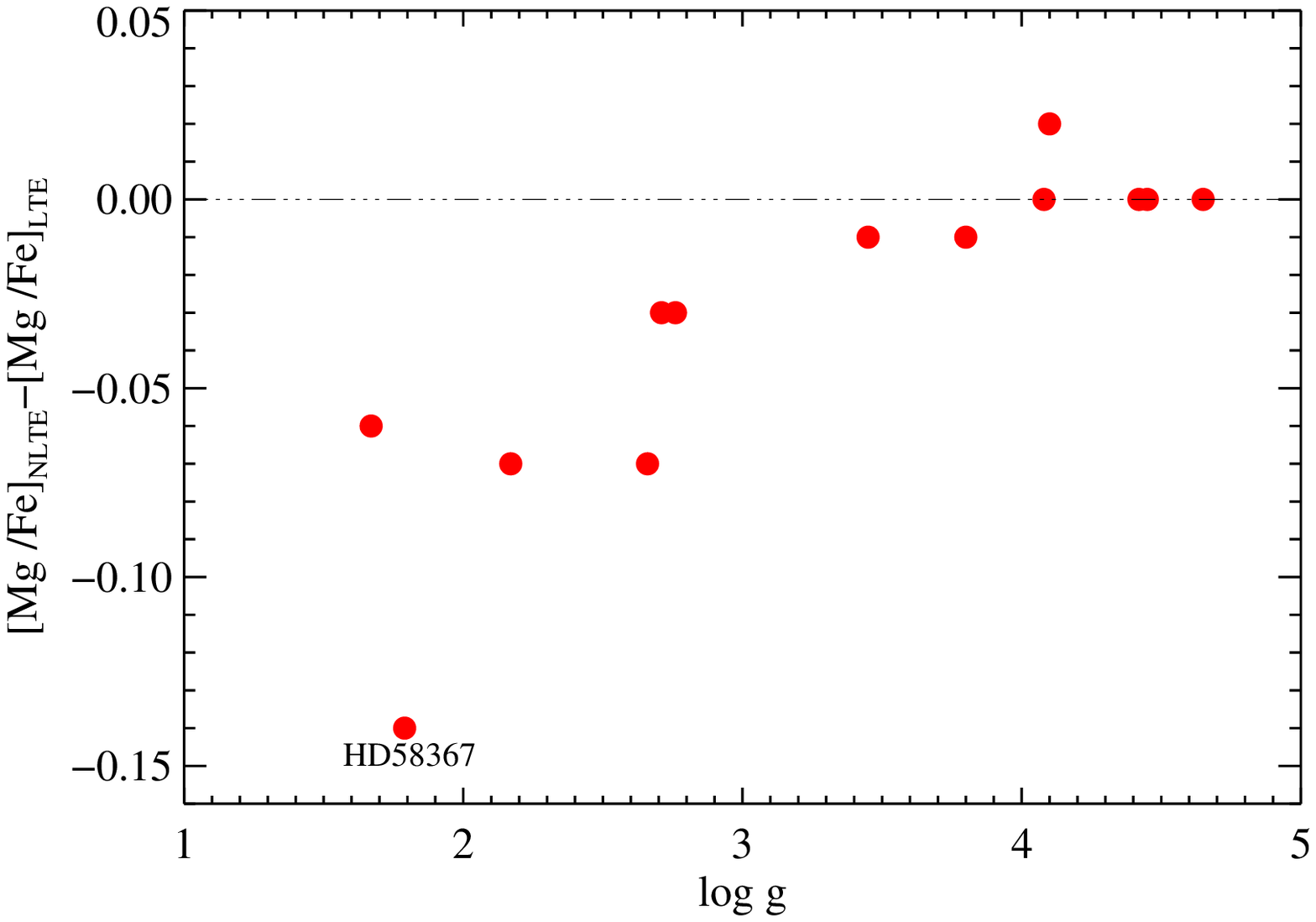}
\caption{The NLTE corrections for Mg 15765\,\AA\ as functions of [Fe/H], $T_{\rm{eff}}$, and log $g$, respectively. (from top to bottom) \label{fig4}}
\end{figure}

As shown in Table\,\ref{tbl-2}, the NLTE effects differ from line to line. Among our four-group {\it{H}}-band lines, the three strong \ion{Mg}{1} lines at 15740, 15748 and 15765\,\AA\ present relatively strong NLTE effects, while the weak one at 15886\,\AA\ shows smaller effects. We plotted the difference of the [Mg/Fe] ratios under NLTE and LTE assumptions for the strong magnesium line at 15765\,\AA\ as a function of metallicity, effective temperature, and surface gravity in Figure \ref{fig4}. It is clear that the NLTE effects depend mainly on surface gravity, similar to our findings for Si (\citetalias{zha16}), namely the corrections increase with decreasing surface gravity, reaching $0.14$\,dex for \object{HD\,58367}. The stellar parameters of this object are $T_{\rm{eff}} =$ 4932\,K, [Fe/H] $= -$0.18\,dex and log $g =$ 1.79\,dex. 

\begin{figure}
\includegraphics[scale=0.45,keepaspectratio=true,clip=true]{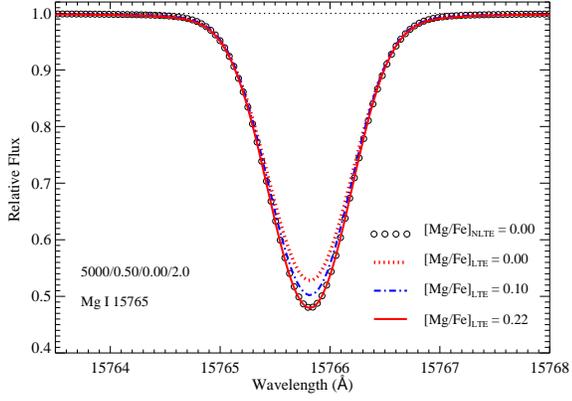}
\caption{LTE and NLTE synthetic spectra of the \ion{Mg}{1} 15765\,\AA\ line with different values of [Mg/Fe] and the same parameters of $T_{\rm{eff}} =$ 5000\,K, log $g = 0.5$\,dex,  [Fe/H] $=$ 0.0\,dex, $\xi_t = 2.0$\,km\,s$^{-1}$. [Mg/Fe] $=$ 0.00, 0.10, 0.22\,dex under LTE respectively, while [Mg/Fe] $=$ 0.00\,dex under NLTE. \label{fig5}}
\end{figure}

It is also interesting to check how large the NLTE corrections can get in the extreme cases for APOGEE. We calculated the NLTE line profile for the \ion{Mg}{1} line at 15765\,\AA\ for a giant with $T_{\rm{eff}} =$ 5000\,K, [Fe/H]=0.0\,dex, log $g$ = 0.5\,dex, [Mg/Fe]=0.0\,dex (see open circle curve in Figure \ref{fig5}). The LTE profile computed with the corresponding NLTE abundance is presented with a dashed curve for comparison. When [Mg/Fe] under LTE and NLTE share the same value of 0.0\,dex, the two profiles are quite different from each other, and the line core in the LTE calculations becomes deeper for increasing values of [Mg/Fe]. The calculated LTE profile obviously deviates from the NLTE one until 
[Mg/Fe] is increased by 0.22\,dex. This indicates that, in this extreme case, the NLTE abundance correction could reach $\sim -$0.22\,dex.

The mean Mg abundances based on the {\it{H}}-band spectra under NLTE and LTE cases, respectively, are listed in Table\,\ref{tbl-3}, along with statistical uncertainties and the mean NLTE corrections. We noted that the value and sign of the NLTE Mg abundance correction are determined by a relative contribution of the core and the wings to the overall line strength. It can be seen that the NLTE corrections range from $-$0.11 to 0.03\,dex.

\begin{figure*}
\plottwo{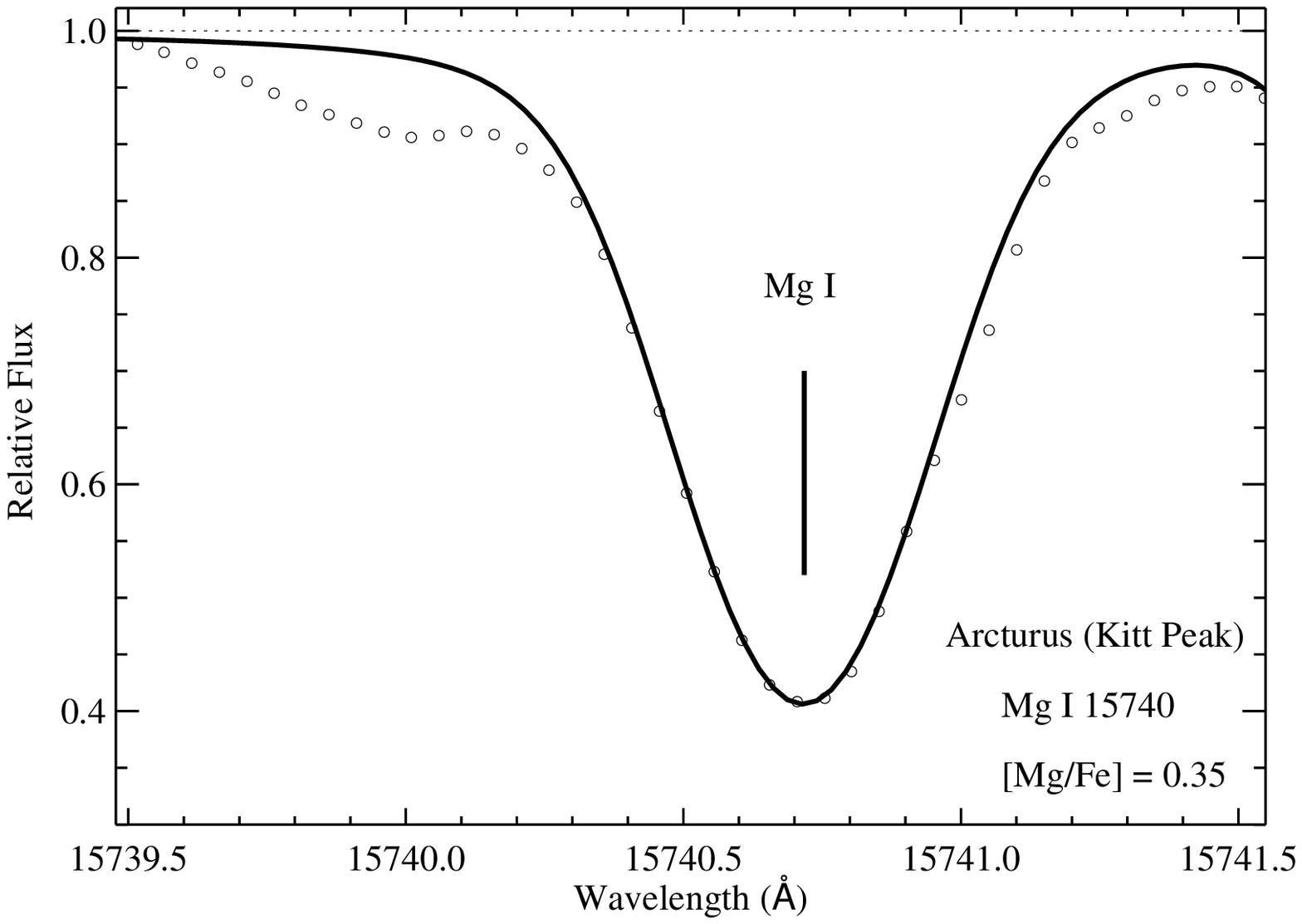}{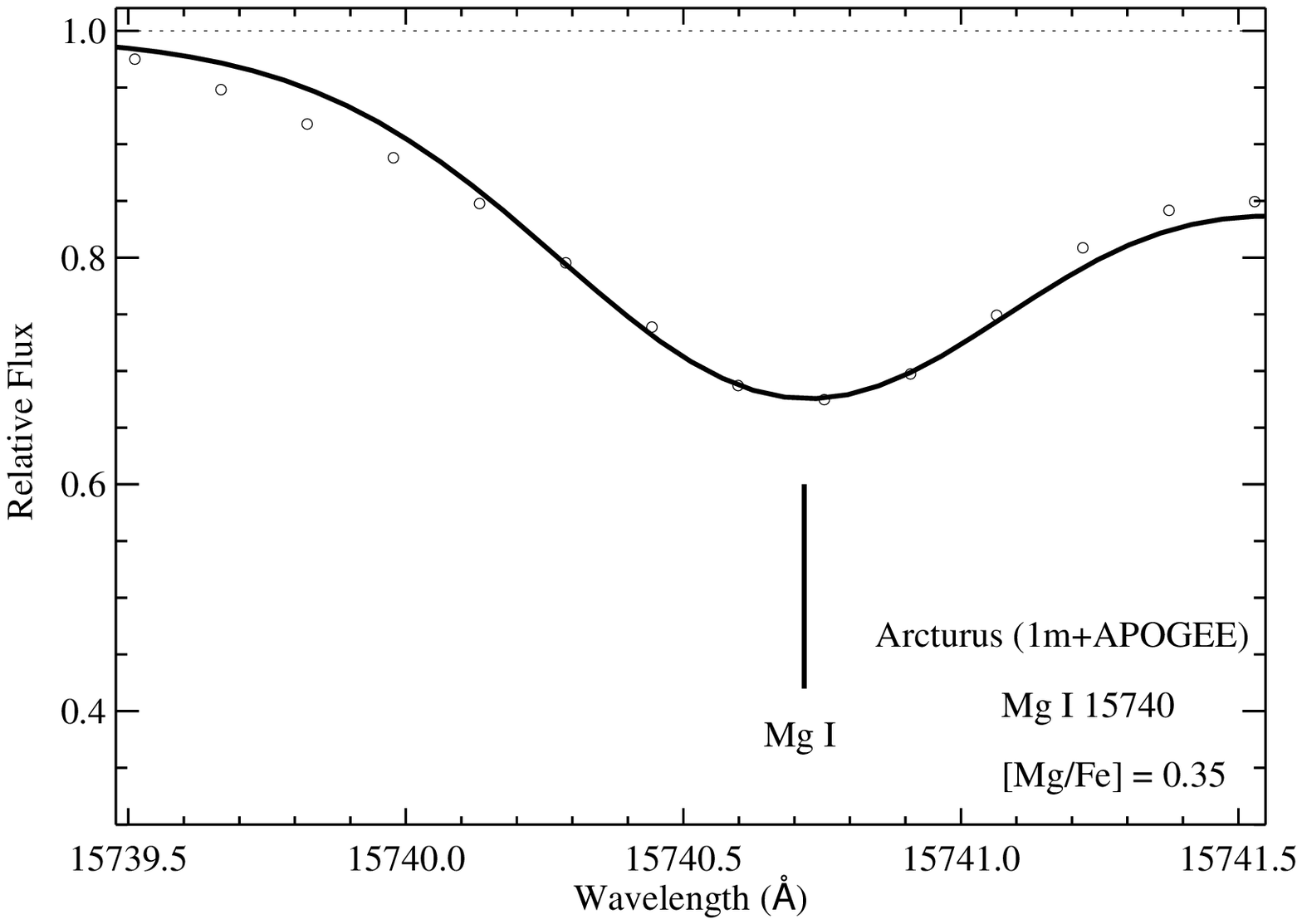}
\plottwo{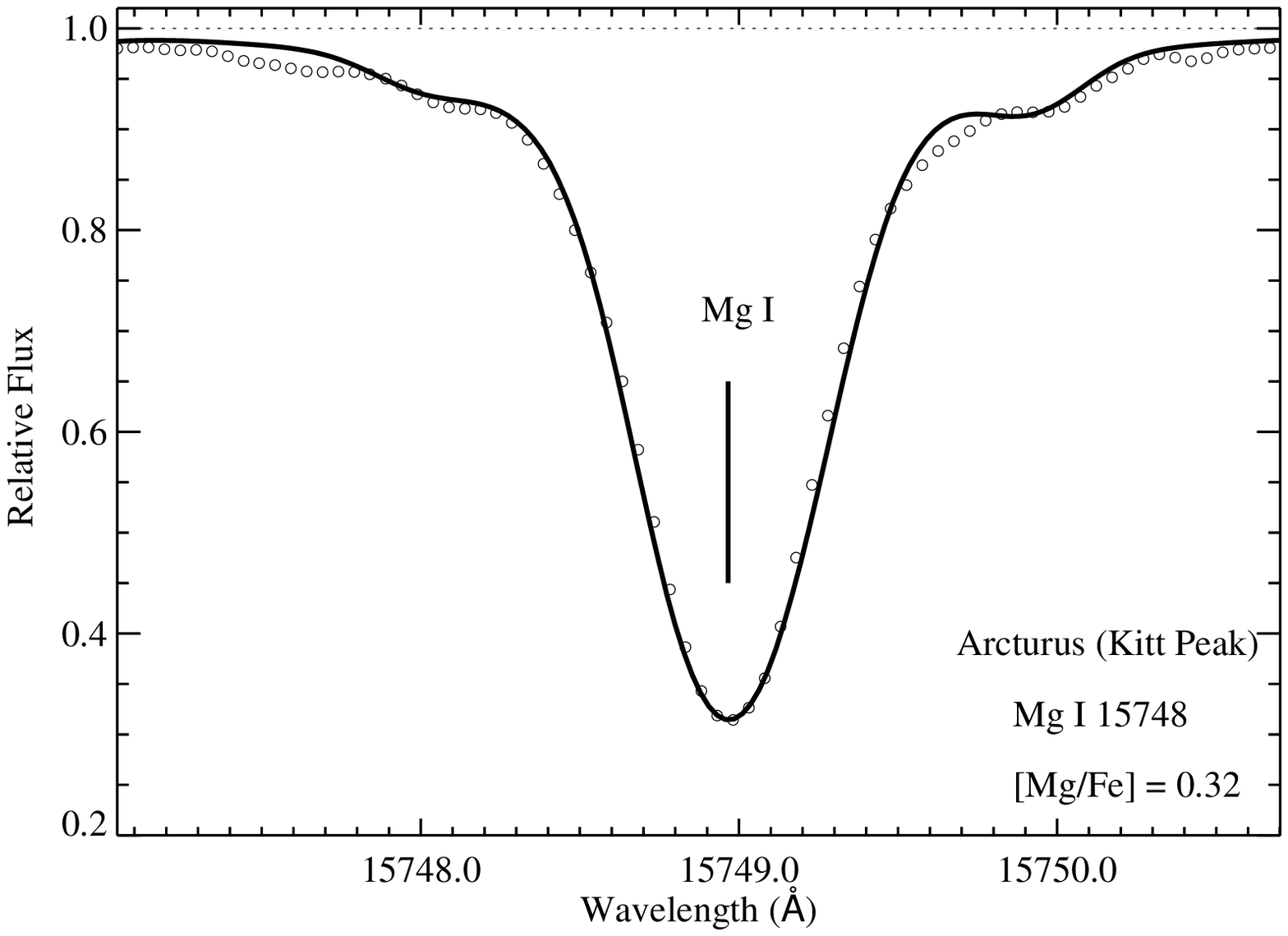}{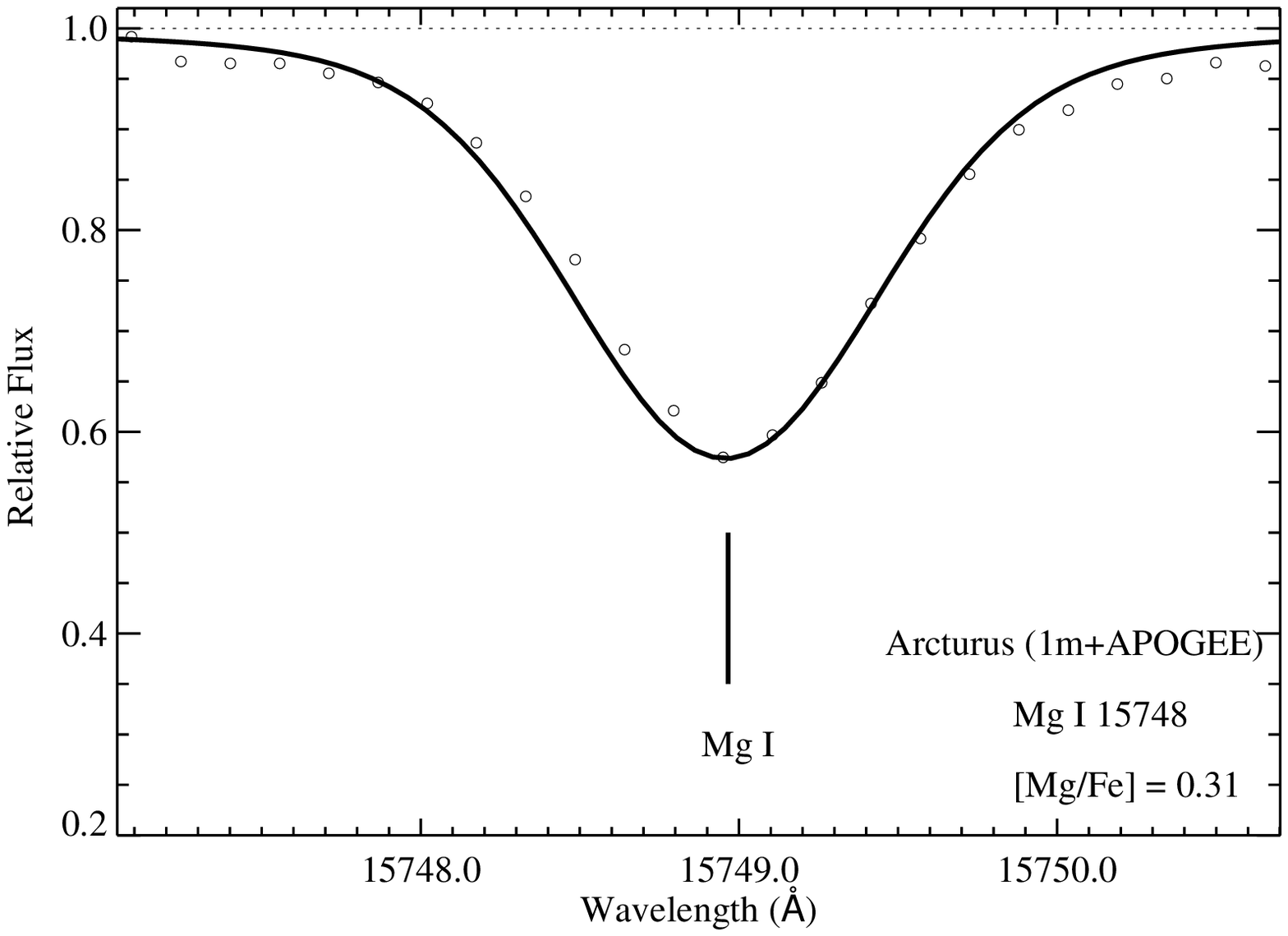}
\plottwo{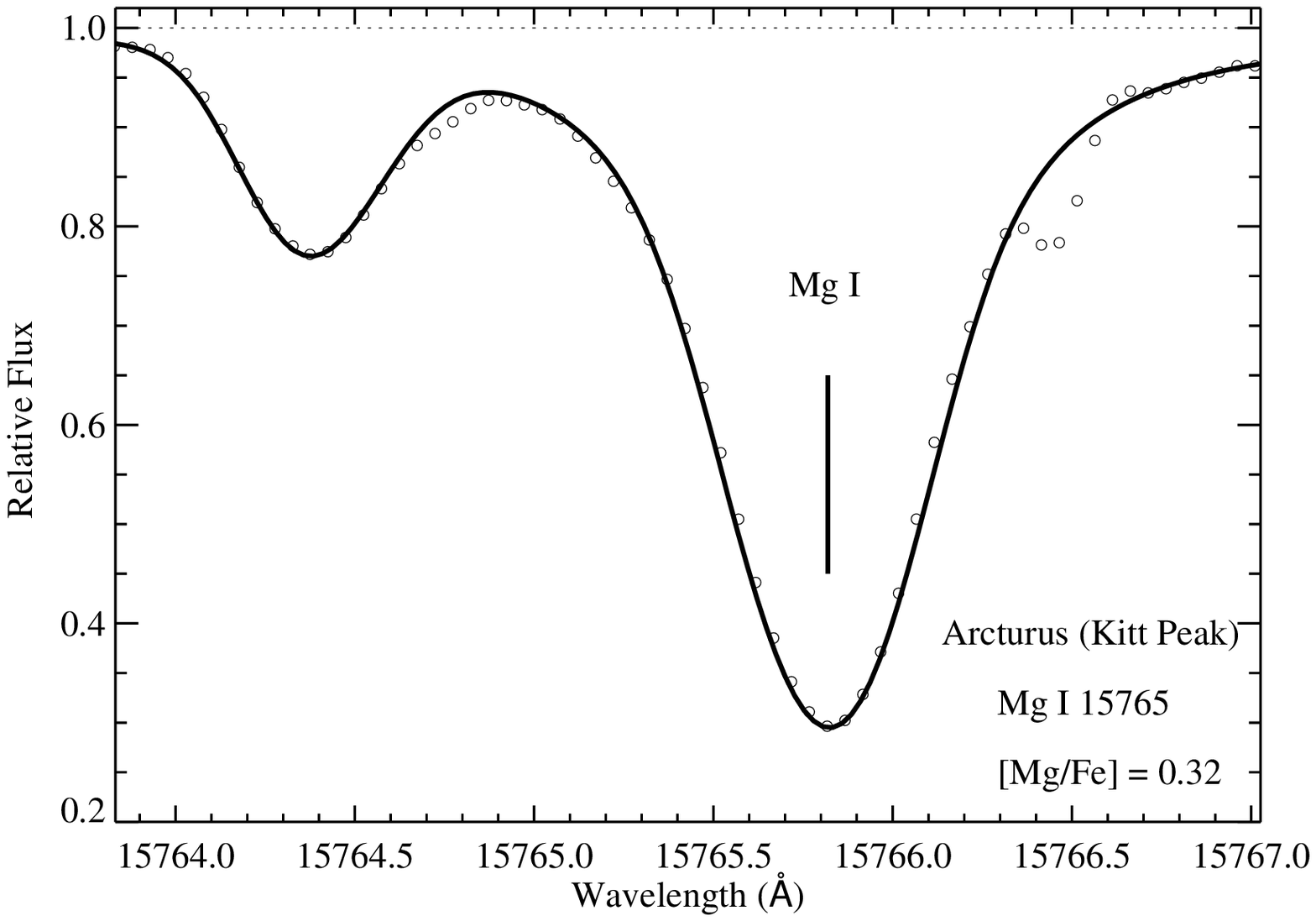}{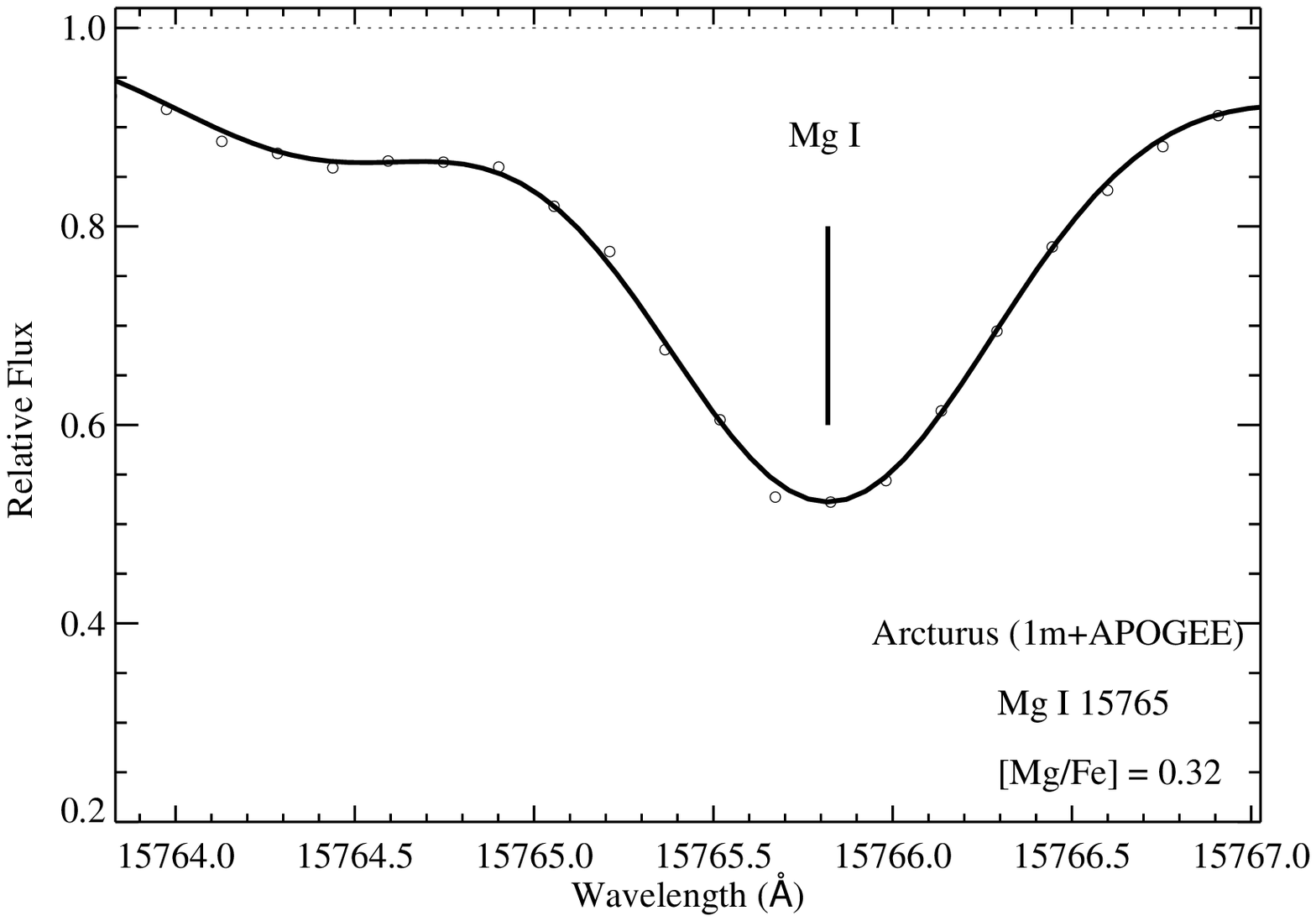}
\plottwo{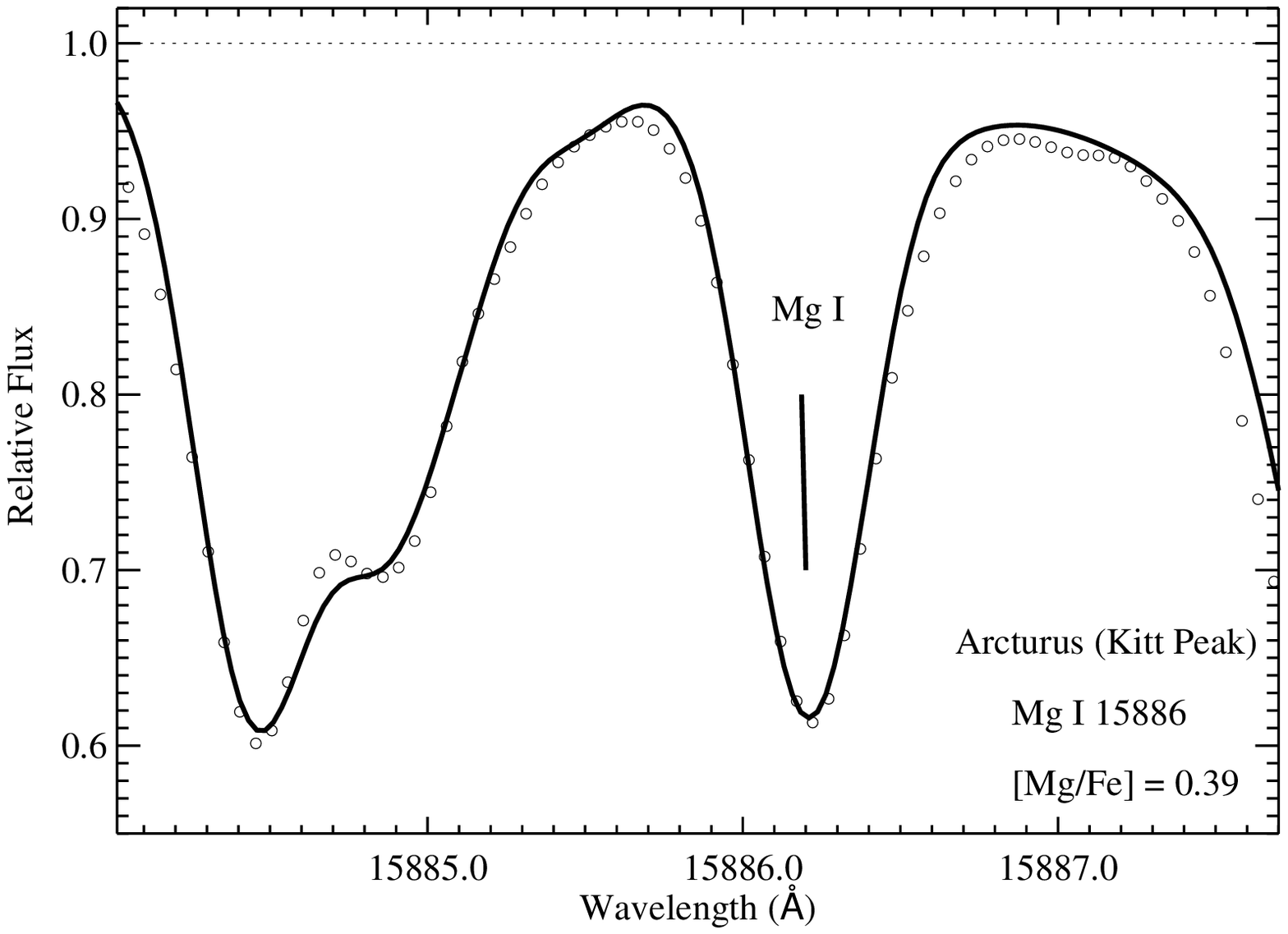}{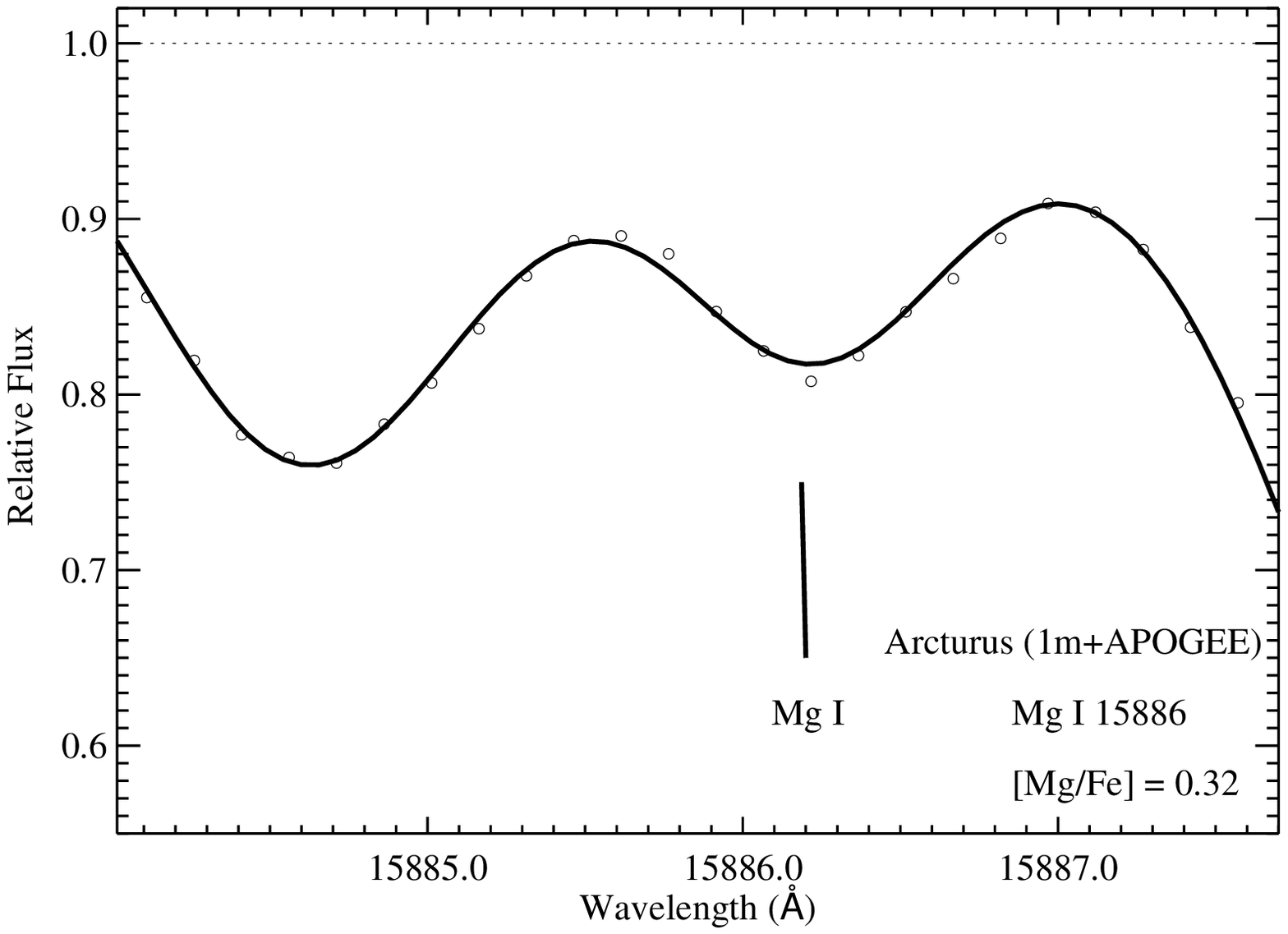}
\caption{The NLTE best fitting profiles (solid line) of the four investigated \ion{Mg}{1} lines for Kitt Peak \citep{hin95} and 1m+APOGEE observed spectra (open circles) of Arcturus. The left column for the spectrum of Arcturus from \citet{hin95} while the right column for the 1m+APOGEE spectrum. \label{fig6}}
\end{figure*}

We derived the Mg abundance of \object{Arcturus} for both the spectrum from \citet{hin95} (R $\sim$ 100,000) and from the 1m$+$ APOGEE (R $\sim$ 22,500), and the best NLTE fitting line profiles are compared with observations in Figure \ref{fig6}. In this figure, the left panel is for the spectrum of Arcturus from Hinkle et al. (1995) and the right panel for the 1m+APOGEE spectrum. We also listed the determined Mg abundance in Table\,\ref{tbl-2} (values for individual lines) and Table\,\ref{tbl-3} (mean values). A consistent mean Mg abundance is derived from these two spectra with a negligible difference of 0.03\,dex.

\begin{deluxetable*}{lrrrrrrrrrrr}
\tablecolumns{12}
\tabletypesize{\scriptsize}
\tablecaption{Magnesium abundances relative to iron under LTE  and NLTE analysis \label{tbl-2}}
\tablewidth{0pc}
\tablehead{
\colhead{} & \multicolumn{2}{c}{15740(\AA)}  & \colhead{} & \multicolumn{2}{c}{15748(\AA)} & \colhead{} & \multicolumn{2}{c}{15765 (\AA)}  & \colhead{} & \multicolumn{2}{c}{15886 (\AA)}\\
\cline{2-3} \cline{5-6} \cline{8-9} \cline{11-12}\\
\colhead{Star} & \colhead{LTE} & \colhead{NLTE} & \colhead{}  & \colhead{LTE} & \colhead{NLTE} & \colhead{} & \colhead{LTE} & \colhead{NLTE} & \colhead{} & \colhead{LTE} & \colhead{NLTE}
}
\startdata
Arcturus\tablenotemark{a}  & 0.39 & 0.35 & & 0.38  &  0.32 & & 0.38 & 0.32 & & 0.37 & 0.39\\
Arcturus\tablenotemark{b}  & 0.40 & 0.35 & & 0.37  &  0.31 & & 0.38 & 0.32 & & 0.31 & 0.32\\
HD 87     &    0.08 &    0.06 & & 0.07 & 0.03 & &    0.06 &    0.03 & & $-$0.04 & $-$0.04\\
HD 6582   &    0.35 &    0.36 & & 0.35 & 0.36 & &    0.35 &    0.35 & &         &        \\
HD 6920   &    0.02 &    0.05 & & 0.06 & 0.05 & &    0.09 &    0.08 & &         &        \\
HD 22675  &    0.08 &    0.07 & &      &      & &    0.06 &    0.03 & &         &        \\
HD 31501  &    0.28 &    0.28 & & 0.21 & 0.21 & &    0.23 &    0.23 & &         &        \\
HD 58367  &    0.23 &    0.16 & & 0.20 & 0.09 & &    0.29 &    0.15 & &         &        \\
HD 67447  &    0.13 &    0.08 & &      &      & &    0.10 &    0.03 & &         &        \\
HD 102870 & $-$0.08 & $-$0.06 & &      &      & & $-$0.10 & $-$0.10 & &         &        \\
HD 103095 &    0.42 &    0.42 & & 0.30 & 0.30 & &    0.31 &    0.31 & &         &        \\
HD 121370 &    0.02 &    0.03 & &      &      & &    0.00 & $-$0.01 & &         &        \\
HD 148816 &    0.32 &    0.35 & & 0.28 & 0.30 & &    0.28 &    0.30 & &         &        \\
HD 177249 &    0.13 &    0.10 & &      &      & &    0.15 &    0.08 & &         &
\enddata
\tablenotetext{a}{The {\it{H}}-band spectrum of Arcturus is from \citet{hin95}.}
\tablenotetext{b}{The {\it{H}}-band spectrum of Arcturus is the 1m $+$ APOGEE one.}
\end{deluxetable*}

\subsubsection{Departures form LTE for the \ion{Mg}{1} Optical lines} \label{nlte_op}
We used six \ion{Mg}{1} optical lines, which are described in Subsection \ref{op_line}, to derive the Mg abundance, and to investigate the NLTE effects for lines in this wavelength band. The mean Mg abundances from the optical spectra under both LTE and NLTE assumptions are separately given in Table\,\ref{tbl-3}. The standard deviation is small, less than 0.07\,dex under NLTE. The [Mg/Fe] ratios for individual lines are also presented in Table\,\ref{tbl-4}.

\subsubsection{Comparisons with the Optical Results and Discussions} \label{com}
\begin{figure}
\includegraphics[scale=0.44,keepaspectratio=true,clip=true]{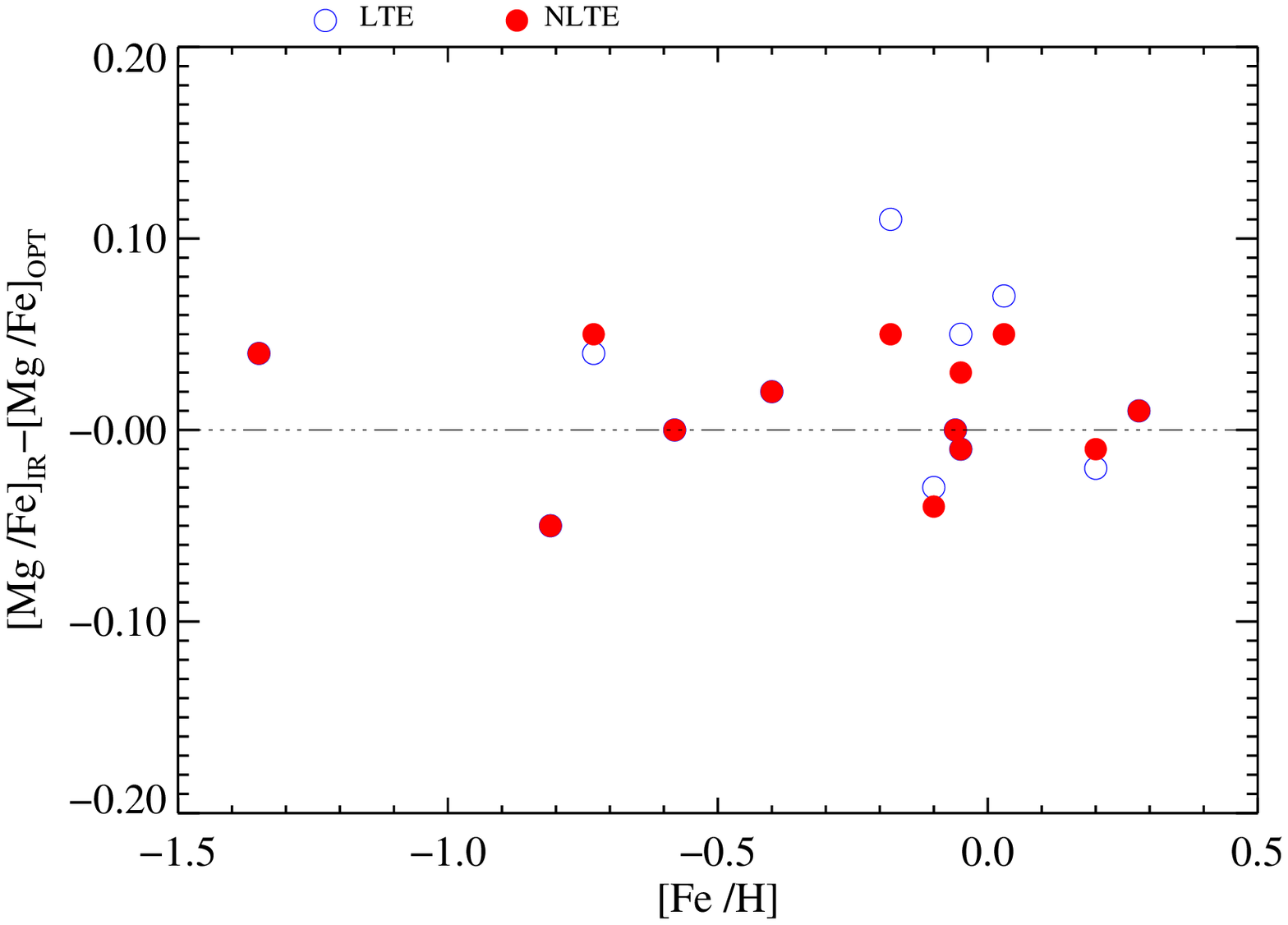}
\caption{The difference between the mean Mg abundances derived from IR and Optical spectra for our sample of stars. \label{fig7}}
\end{figure}

We present the average Mg abundances based on the IR and optical spectra under the LTE and NLTE assumptions, respectively, as well as the stellar parameters adopted for our sample stars in Table\,\ref{tbl-3}. In Figure \ref{fig7}, the differences of the mean Mg abundances between {\it{H}}-band and optical lines are plotted against the metallicity for our target stars, where the open circles represent the LTE abundances, while filled circles stand for NLTE results. It can be clearly seen a better consistent result can be derived when the NLTE effects considered, although the differences are small ($\sim$ 0.1\,dex) both in LTE and NLTE.  

\begin{deluxetable*}{lrcrrrrrrrr}
\tabletypesize{\scriptsize}
\tablecaption{Stellar magnesium LTE and NLTE abundances\label{tbl-3}}
\tablewidth{0pt}
\tablehead{
\colhead{Star} & \colhead{$T_{\rm{eff}}$} & \colhead{log $g$} & \colhead{[Fe/H]} & \colhead{$\xi_t$} & \colhead{[{\ion{Mg}{1}}$_{LTE}$/Fe](IR)} & \colhead{[{\ion{Mg}{1}}$_{NLTE}$/Fe](IR)} & \colhead{$\Delta_{\rm{IR}}$} & \colhead{[\ion{Mg}{1}$_{LTE}$/Fe](OPT)} & \colhead{[{\ion{Mg}{1}}$_{NLTE}$/Fe](OPT)} & \colhead{$\Delta_{\rm{OPT}}$}
}
\startdata
Arcturus\tablenotemark{a} & 4275 & 1.67 & $-$0.58 & 1.60 &    0.38$\pm$0.01 &    0.35$\pm$0.03 & $-$0.03 &    0.38$\pm$0.07 &    0.35$\pm$0.03 & $-$0.03\\
Arcturus\tablenotemark{b} & 4275 & 1.67 & $-$0.58 & 1.60 &    0.36$\pm$0.04 &    0.32$\pm$0.02 & $-$0.04 &                                     &        \\
HD 87                     & 5053 & 2.71 & $-$0.10 & 1.35 &    0.04$\pm$0.06 &    0.02$\pm$0.04 & $-$0.02 &    0.07$\pm$0.06 &    0.06$\pm$0.03 & $-$0.01\\
HD 6582                   & 5390 & 4.42 & $-$0.81 & 0.90 &    0.35$\pm$0.00 &    0.36$\pm$0.01 &    0.01 &    0.40$\pm$0.02 &    0.41$\pm$0.02 &    0.01\\
HD 6920                   & 5845 & 3.45 & $-$0.06 & 1.40 &    0.06$\pm$0.04 &    0.06$\pm$0.02 &    0.00 &    0.06$\pm$0.06 &    0.06$\pm$0.07 &    0.00\\
HD 22675                  & 4901 & 2.76 & $-$0.05 & 1.30 &    0.07$\pm$0.01 &    0.05$\pm$0.03 & $-$0.02 &    0.08$\pm$0.05 &    0.06$\pm$0.03 & $-$0.02\\
HD 31501\tablenotemark{c} & 5320 & 4.45 & $-$0.40 & 1.00 &    0.24$\pm$0.04 &    0.24$\pm$0.04 &    0.00 &    0.22$\pm$0.00 &    0.22$\pm$0.00 &    0.00\\
HD 58367                  & 4932 & 1.79 & $-$0.18 & 2.00 &    0.24$\pm$0.05 &    0.13$\pm$0.04 & $-$0.11 &    0.13$\pm$0.10 &    0.08$\pm$0.03 & $-$0.05\\
HD 67447                  & 4933 & 2.17 & $-$0.05 & 2.12 &    0.12$\pm$0.02 &    0.06$\pm$0.04 & $-$0.06 &    0.07$\pm$0.04 &    0.02$\pm$0.01 & $-$0.05\\
HD 102870                 & 6070 & 4.08 &    0.20 & 1.20 & $-$0.09$\pm$0.01 & $-$0.08$\pm$0.03 &    0.01 & $-$0.08$\pm$0.06 & $-$0.07$\pm$0.06 &    0.01\\
HD 103095                 & 5085 & 4.65 & $-$1.35 & 0.80 &    0.34$\pm$0.07 &    0.34$\pm$0.07 &    0.00 &    0.30$\pm$0.05 &    0.30$\pm$0.05 &    0.00\\
HD 121370                 & 6020 & 3.80 &    0.28 & 1.40 &    0.01$\pm$0.01 &    0.01$\pm$0.03 &    0.00 &    0.00$\pm$0.04 &    0.00$\pm$0.05 &    0.00\\
HD 148816                 & 5830 & 4.10 & $-$0.73 & 1.40 &    0.29$\pm$0.02 &    0.32$\pm$0.03 &    0.03 &    0.25$\pm$0.05 &    0.27$\pm$0.05 &    0.02\\
HD 177249                 & 5273 & 2.66 &    0.03 & 1.65 &    0.14$\pm$0.01 &    0.09$\pm$0.01 & $-$0.05 &    0.07$\pm$0.10 &    0.04$\pm$0.05 & $-$0.03
\enddata
\tablecomments{$\rm\Delta_{ir}$ and $\rm\Delta_{opt}$ stand for the NLTE effects ($\Delta$ $=$ $\rm{log}\,\varepsilon_{NLTE}$ $-$ $\rm{log}\,\varepsilon_{LTE}$) derived from IR and optical spectra respectively.}
\tablenotetext{a}{The H-band spectrum of Arcturus is from \citet{hin95}.}
\tablenotetext{b}{The H-band spectrum of Arcturus is the 1m $+$ APOGEE one.}
\tablenotetext{c}{Only one line is calculated, so the error is zero.}
\end{deluxetable*}

\section{CONCLUSIONS} \label{con}
We verified the reliability of our Mg atomic model for the {\it{H}}-band line formation based on the high quality {\it{H}}-band spectra, and investigated NLTE effects on \ion{Mg}{1} {\it{H}}-band lines for 13 FGK sample stars. 
A detailed analysis based on a line-to-line differential analysis relative to the Sun both under LTE and NLTE allows us to obtain accurate Mg abundances. Our conclusions can be summarized as follows.
\begin{itemize}
\item The mean Mg abundance differences between the {\it{H}}-band and optical lines are within 0.05\,dex for all our sample stars when NLTE effects are included, which suggests that our Mg atomic model can be applied to study the formation of {\it{H}}-band \ion{Mg}{1} lines.
\item It is shown that the NLTE effects tend to be large for strong \ion{Mg}{1} lines at 15740, 15748 and 15765\,\AA, and smaller for the relatively weak line at 15886\,\AA. The NLTE correction is as large as $\sim -$0.14\,dex for the strong \ion{Mg}{1} line in our sample stars, and for the extreme case in APOGEE it reaches up to $\sim -$0.22\,dex. Thus, departures from LTE need to be considered in the Mg abundance analysis, especially when only strong lines are available.
\item The NLTE effects in {\it{H}}-band Mg lines are very sensitive to surface gravity, increasing for lower gravities. Therefore, they need to be considered for APOGEE since most of their sample are giant, RGB stars. The corrections are always negative for cool stars, and therefore LTE Mg abundances will be overestimated.
\end{itemize}

We conclude that, it is important to consider NLTE effects in the calculations of {\it{H}}-band Mg lines.

\acknowledgments
This research is supported by National Key Basic Research Program of China 2014CB845700, and by the National Natural Science Foundation of China under grant Nos. 11321064, 11233004, 11390371, 11473033, 11428308, U1331122. CAP is thankful to the Spanish MINECO for support through grant AYA2014-56359-P.

We thank Dr. Takeda Y., Dr. Sato B. and Dr. Liu Y. J. for providing us the optical data. We acknowledge the support of the staff at the Xinglong 2.16m telescope.

NSO/Kitt Peak FTS data used here were produced by NSF/NOAO.

Funding for SDSS-III has been provided by the Alfred P. Sloan Foundation, the Participating Institutions, the National Science Foundation, and the U.S. Department of Energy Office of Science. The SDSS-III web site is \texttt{http://www.sdss3.org/}.

SDSS-III is managed by the Astrophysical Research Consortium for the Participating Institutions of the SDSS-III Collaboration including the University of Arizona, the Brazilian Participation Group, Brookhaven National Laboratory, Carnegie Mellon University, University of Florida, the French Participation Group, the German Participation Group, Harvard University, the Instituto de Astrofisica de Canarias, the Michigan State/Notre Dame/JINA Participation Group, Johns Hopkins University, Lawrence Berkeley National Laboratory, Max Planck Institute for Astrophysics, Max Planck Institute for Extraterrestrial Physics, New Mexico State University, New York University, Ohio State University, Pennsylvania State University, University of Portsmouth, Princeton University, the Spanish Participation Group, University of Tokyo, University of Utah, Vanderbilt University, University of Virginia, University of Washington, and Yale University.


\clearpage
\appendix
\clearpage

\begin{deluxetable}{lrrrrrrrrrrrrrrrrrrr}
\tablecolumns{13}
\tabletypesize{\scriptsize}
\tablecaption{Magnesium abundances relative to iron based on optical \ion{Mg}{1} lines under LTE and NLTE analysis \label{tbl-4}}
\tablewidth{0pt}
\tablehead{
\colhead{} & \multicolumn{2}{c}{4571 (\AA)}  & \multicolumn{2}{c}{4702 (\AA)} & \multicolumn{2}{c}{5172 (\AA)} & \multicolumn{2}{c}{5183 (\AA)} & \multicolumn{2}{c}{5528 (\AA)} & \multicolumn{2}{c}{5711 (\AA)} \\
\cline{2-3} \cline{4-5} \cline{6-7} \cline{8-9} \cline{10-11} \cline{12-13} \\
\colhead{Star} & \colhead{LTE} & \colhead{NLTE} & \colhead{LTE} & \colhead{NLTE} & \colhead{LTE} & \colhead{NLTE}& \colhead{LTE} & \colhead{NLTE} & \colhead{LTE} & \colhead{NLTE} & \colhead{LTE} & \colhead{NLTE}
}
\startdata
Arcturus   &    0.37 &    0.38 &      &      &    0.32 &    0.33 &    0.30 &    0.31 &    0.43 &    0.38 &    0.46 &    0.34 \\
HD\,87     &         &         &      &      &    0.02 &    0.04 &    0.02 &    0.03 &    0.13 &    0.08 &    0.11 &    0.08 \\
HD\,6582   &         &         & 0.42 & 0.42 &         &         &         &         &    0.39 &    0.38 &    0.40 &    0.42 \\
HD\,6920   &         &         & 0.15 & 0.16 & $-$0.01 &    0.00 &    0.02 &    0.03 &    0.06 &    0.02 &    0.06 &    0.10 \\
HD\,22675  &         &         &      &      &    0.05 &    0.05 &    0.03 &    0.04 &    0.11 &    0.07 &    0.14 &    0.10 \\
HD\,31501  &         &         &      &      &         &         &         &         &         &         &    0.22 &    0.22 \\
HD\,58367  &         &         &      &      &    0.06 &    0.08 &    0.04 &    0.06 &    0.25 &    0.12 &    0.15 &    0.08 \\
HD\,67447  &         &         & 0.03 & 0.03 &         &         &         &         &    0.11 &    0.01 &    0.07 &    0.02 \\
HD\,102870 & $-$0.13 & $-$0.10 & 0.02 & 0.03 & $-$0.11 & $-$0.11 & $-$0.13 & $-$0.13 & $-$0.06 & $-$0.09 & $-$0.04 & $-$0.01 \\
HD\,103095 &    0.22 &    0.22 & 0.32 & 0.32 &    0.32 &    0.32 &    0.30 &    0.30 &    0.27 &    0.26 &    0.36 &    0.37 \\
HD\,121370 &         &         &      &      & $-$0.03 & $-$0.02 & $-$0.03 & $-$0.03 &    0.00 & $-$0.02 &    0.06 &    0.08 \\
HD\,148816 &    0.21 &    0.24 & 0.33 & 0.35 &    0.23 &    0.24 &    0.19 &    0.21 &    0.30 &    0.30 &    0.26 &    0.31 \\
HD\,177249 &         &         &      &      &    0.00 &    0.01 &    0.00 &    0.01 &    0.21 &    0.11 &    0.07 &    0.04
\enddata
\end{deluxetable}

\end{document}